\documentclass[a4paper,11pt]{article}
\usepackage{pos}
\usepackage{graphicx}
\usepackage[normalem]{ulem}

\title{Dark photon in parity-violating electron scatterings}
\author[a]{A.~W.~Thomas}
\author*[a]{X.~G.~Wang}
\author[a]{A.~G.~Williams}

\affiliation[a]{CSSM and ARC Centre of Excellence for Dark Matter Particle Physics,\\ 
Department of Physics, University of Adelaide, Adelaide, SA 5005, Australia
}

\emailAdd{xuan-gong.wang@adelaide.edu.au}

\abstract{We proposed that parity-violating electron scattering (PVES) offers a powerful tool to probe the hypothetical dark photon. 
We calculated the dark photon contributions to PVES asymmetries in both elastic and deep-inelastic scattering (DIS).
These contributions  are characterised  by the corrections to the standard model couplings 
$C_{1q}, \, C_{2q}$, and $C_{3q}$. At low scales, the corrections to $C_{1q}$ and $C_{3q}$ could be as large as $5\%$ were a dark photon to exist. In DIS at very high $Q^2$, of relevance to HERA or the EIC,  the dark photon could induce substantial corrections to $C_{2q}$, suggesting as large as $10\%$ uncertainties in the extraction of valence parton distribution functions. 
 We also extracted the favoured regions of the dark photon parameter space by fitting the parity violation data and the CDF $W$ boson mass, which prefer a heavy dark photon with mass above the $Z$-boson mass.
 }

\FullConference{The XVIth Quark Confinement and the Hadron Spectrum Conference (QCHSC24)\\
 19-24 August, 2024\\
 Cairns Convention Centre, Cairns, Queensland, Australia\\}

\begin{document}
\maketitle

\section{Introduction}

In searching for new physics beyond the Standard Model (SM), extensions in the gauge sector of the electroweak theory have received increasingly interest. Extra $U(1)$ gauge fields could be introduced, either through kinetic mixing in the dark photon model~\cite{Fayet:1980ad, Fayet:1980rr, Holdom:1985ag, Okun:1982xi} or anomaly-free $U(1)'$ charges in the $Z'$ models~\cite{Fayet:1990wx, Leike:1998wr}. These models are also appealing as portals connecting to the dark matter sector. In this contribution, we will focus on the dark photon hypothesis.

The dark photon has been searched for in fixed target experiments~\cite{NA64:2022yly}, and at the electron~\cite{BaBar:2014zli, BaBar:2017tiz} and hadron colliders~\cite{LHCb:2019vmc, CMS:2019buh}.
So far, there has been no direct evidence for its existence.
Rather stringent constraints have been placed on the kinetic mixing parameter, leading to an upper limit of $\epsilon \le 10^{-3}$ for dark photon masses below 200 GeV. However, these limits could be significantly relaxed in light of the potential couplings of the dark boson to dark matter particles~\cite{Abdullahi:2023tyk}.

Theoretical constraints on the dark photon parameters have also been derived from the measurements of $g-2$ for the muon~\cite{Pospelov:2008zw, Davoudiasl:2012qa}, $e-p$ deep inelastic scattering~\cite{Kribs:2020vyk, Thomas:2021lub, Hunt-Smith:2023sdz}, electroweak precision observables~\cite{Hook:2010tw, Curtin:2014cca, Loizos:2023xbj}, and rare decays of kaon and $B$ mesons~\cite{Davoudiasl:2012ag, Datta:2022zng, Wang:2023css}. In particular, a recent global QCD analysis of electron-nucleon deep inelastic scattering and related high-energy data within the JAM framework found a significant reduction in the $\chi^2$ by including the dark photon. This provides the first hint of the existence of the dark photon, although indirectly~\cite{Hunt-Smith:2023sdz}. In contrast, no improvement in the $\chi^2$ was found if instead we included the $U(1)_{B-L}$ $Z'$ boson in such a global fit analysis~\cite{Wang:2024gvt}.

In this contribution, we report a new proposal for dark photon searches through parity-violating electron scattering (PVES) experiments~\cite{Thomas:2022qhj,Thomas:2022gib}. In Section~\ref{sec:PVES}, we show the PVES asymmetries in both elastic and deep-inelastic scatterings. The dark photon formalism is given in Section~\ref{sec:darkphoton}. We present the sensitivity of the beam asymmetries to the dark photon parameters in Section~\ref{sec:sensitivity}, and the favoured dark photon parameters from fits to parity violation data and the CDF $W$ boson mass in Section~\ref{sec:fit}. Our concluding remarks are given in Section~\ref{sec:conclusion}.

\section{Parity-violating electron scattering}
\label{sec:PVES}
In the scattering of longitudinally polarised electrons on an unpolarised target, the parity-violation effect is characterised by the asymmetry between left- and right-handed electrons,
\begin{equation}
\label{eq:A_PV}
A_{\rm PV} = \frac{\sigma_R - \sigma_L}{\sigma_R + \sigma_L} \, ,
\end{equation}
where $\sigma_{R,L} = d^2 \sigma_{R,L}/d\Omega dE'$ are the double differential cross sections of right-handed (R) and left-handed (L) electrons, respectively.

For elastic scattering, the asymmetry can be expressed in terms of the weak form factor $F_W$ and the charge form factor $F_C$~\cite{Horowitz:1999fk}, 
\begin{equation}
A^{\rm el}_{\rm PV} = \frac{G_F Q^2 |Q^{(W)}_{N,Z}|}{4\sqrt{2} \pi \alpha Z} \frac{F_W(Q^2)}{F_C(Q^2)}\, ,
\end{equation}
where $G_F = 1.1663787\times 10^{-5} {\rm GeV}^{-2}$ is the Fermi constant. $Q^{(W)}_{N,Z}$ is the weak charge of the target nucleus with $N$ neutrons and $Z$ protons,
which reads at tree level
\begin{equation}
Q^{(W)}_{N,Z} = - 2 \Big[ (2 C_{1u} + C_{1d}) Z + (C_{1u} + 2 C_{1d}) N \Big]\, .
\end{equation}

In the case of deep inelastic scattering (DIS), especially from a deuteron target, the PVES asymmetry and the lepton charge asymmetry provide direct connection to the fundamental weak couplings~\cite{Zheng:2021hcf}
\begin{eqnarray}
\label{eq:A_d}
A_d^{e^-_R - e^-_L} &=& \frac{3 G_F Q^2}{10 \sqrt{2} \pi \alpha_{em}} \Big[ (2 C_{1u} - C_{1d}) + R_V Y (2 C_{2u} - C_{2d}) \Big]\, ,\nonumber\\
A_d^{e^+ - e^-} &=& -  \frac{3 G_F Q^2 Y}{2 \sqrt{2} \pi \alpha_{em}} \frac{R_V (2 C_{3u} - C_{3d})}{5 + 4 R_C + R_S}\, ,
\end{eqnarray}
where $C_{1q}$, $C_{2q}$ and $C_{3q}$ are the products of weak couplings to the electron and quarks. In the SM,
\begin{equation}
\label{eq:C_q}
C_{1q}^{\rm SM} = 2 g^e_A g^q_V\, ,\ \ C_{2q}^{\rm SM} = 2 g^e_V g^q_A\, ,\ \ C_{3q}^{\rm SM} = - 2 g^e_A g^q_A\, ,
\end{equation}
where the tree-level couplings are
\begin{eqnarray}
\label{eq:couplings}
\{ g^e_V, g^u_V, g^d_V\} &=& \{ - \frac{1}{2} + 2 \sin^2\theta_W\, , \frac{1}{2} - \frac{4}{3}\sin^2\theta_W \, , - \frac{1}{2} + \frac{2}{3}\sin^2\theta_W\}\, , \nonumber\\
\{ g^e_A, g^u_A, g^d_A\} &=& \{ - \frac{1}{2}\, , \frac{1}{2}\, , -\frac{1}{2} \} \, ,
\end{eqnarray}
with $\theta_W$ the Weinberg angle. 
Possible corrections to the asymmetries in Eq.~(\ref{eq:A_d}) arising from charge symmetry violation, strange and charm quark distributions are discussed in Ref.~\cite{Wang:2024mll}.

PVES offers direct and precise measurements on $C_{1q}$, $C_{2q}$ and $C_{3q}$. Any deviations from their SM predictions would imply signals of new physics.

\section{Dark photon formalism}
\label{sec:darkphoton}
The dark photon is usually introduced as an extra $U(1)$ gauge boson~\cite{Fayet:1980ad, Fayet:1980rr, Holdom:1985ag}, 
interacting with SM particles through kinetic mixing with hypercharge~\cite{Okun:1982xi}
\begin{eqnarray}
\label{eq:L}
{\cal L} & \supset & 
- \frac{1}{4} F'_{\mu\nu} F'^{\mu\nu} + \frac{1}{2} m^2_{A'} A'_{\mu} A'^{\mu} 
+ \frac{\epsilon}{2 \cos\theta_W} F'_{\mu\nu} B^{\mu\nu} 
\, ,
\end{eqnarray}
where $F'_{\mu\nu}$ is the dark photon strength tensor and $\epsilon$ is the mixing parameter. We use $A'$ and $\bar{Z}$ to denote the unmixed versions of the dark photon and the SM neutral weak boson, respectively. 

By diagonalising the mass-squared matrix, one can define the physical $Z$ and $A_D$ with masses~\cite{Kribs:2020vyk}
\begin{eqnarray}
\label{eq:m_Z_AD}
m^2_{Z, A_D} &=& \frac{m_{\bar{Z}}^2}{2} [ 1 + \epsilon_W^2 + \rho^2
\pm {\rm sign}(1-\rho^2) \sqrt{(1 + \epsilon_W^2 + \rho^2)^2 - 4 \rho^2} ] \, ,
\end{eqnarray}
where $\alpha$ is the $\bar{Z}-A'$ mixing angle,
\begin{eqnarray}
\tan \alpha &=& \frac{1}{2\epsilon_W} \Big[ 1 - \epsilon^2_W - \rho^2 - {\rm sign}(1-\rho^2) \sqrt{4\epsilon_W^2 + ( 1 - \epsilon_W^2 - \rho^2)^2} \Big] \, ,
\end{eqnarray}
with
\begin{eqnarray}
\label{eq:epsW-rho}
\epsilon_W &=& \frac{\epsilon \tan \theta_W}{\sqrt{1 - \epsilon^2/\cos^2\theta_W}} ,\nonumber\\
\rho &=& \frac{m_{A'}/m_{\bar{Z}}}{\sqrt{1 - \epsilon^2/\cos^2\theta_W}} \, .
\end{eqnarray}

The lowest order SM couplings of the $Z$ boson to leptons and quarks, $C_{\bar{Z}}^v = \{g^e_V, g^u_V, g^d_V\}$ and $C_{\bar{Z}}^a = \{g^e_A, g^u_A, g^d_A\}$, will be shifted to~\cite{Kribs:2020vyk,Thomas:2022qhj}
\begin{eqnarray}
\label{eq:C_Z}
C_Z^v &=& (\cos\alpha - \epsilon_W \sin\alpha) C_{\bar{Z}}^v + 2 \epsilon_W \sin\alpha \cos^2 \theta_W C_{\gamma}^v ,\nonumber\\
C_Z^a &=& (\cos\alpha - \epsilon_W \sin\alpha) C_{\bar{Z}}^a\, ,
\end{eqnarray}
where 
$C_{\gamma}^v = \{C^e_{\gamma}, C^u_{\gamma}, C^d_{\gamma}\} = \{-1, 2/3, - 1/3 \}$. 
Likewise, the couplings of the physical dark photon $A_D$ to SM fermions are given by
\begin{eqnarray}
\label{eq:C_AD}
C_{A_D}^v &=& - (\sin\alpha + \epsilon_W \cos\alpha) C_{\bar{Z}}^v + 2 \epsilon_W \cos\alpha \cos^2 \theta_W C_{\gamma}^v ,\nonumber\\
C_{A_D}^a &=& - (\sin\alpha + \epsilon_W \cos\alpha) C_{\bar{Z}}^a 
\, .
\end{eqnarray}
Due to its nonzero axial-vector couplings, the dark photon will also contribute to the parity-violating electron scatterings.

\section{PVES sensitivity to dark photon}
\label{sec:sensitivity}

The double differential cross section with the dark photon contributions can be expressed as~\cite{Thomas:2022qhj}
{\small
\begin{eqnarray}
\frac{d^2 \sigma}{dx dy} 
&=& \frac{4\pi \alpha^2 s}{Q^4}
\Big(
[x y^2 F_1^{\gamma} + f_1(x,y) F_2^{\gamma}] \nonumber\\
&& - \frac{1}{\sin^2 2\theta_W}\frac{Q^2}{Q^2 + M_Z^2} (C_{Z,e}^v - \lambda C_{Z,e}^a)
 [x y^2 F_1^{\gamma Z} + f_1(x,y) F_2^{\gamma Z} - \lambda x y (1-\frac{y}{2}) F_3^{\gamma Z}]  \nonumber\\
&&
-  \frac{1}{\sin^2 2\theta_W} \frac{Q^2}{Q^2 + M_{A_D}^2} (C_{A_D,e}^v - \lambda C_{A_D,e}^a)
[x y^2 F_1^{\gamma A_D} + f_1(x,y) F_2^{\gamma A_D} - \lambda x y (1-\frac{y}{2}) F_3^{\gamma A_D}]
\Big) \, ,\nonumber\\
\end{eqnarray}
}
where  $f_1(x,y) = 1 - y - xyM/2E$ and $\lambda = + 1 (-1)$ represents positive (negative) initial electron helicity. The cross sections for positron scattering can be obtained with $C_{Z,e}^a$ and $C_{A_D,e}^a$ being replaced by $-C_{Z,e}^a$ and $-C_{A_D,e}^a$, respectively~\cite{Anselmino:1993tc}.
 
The numerator in Eq.~(\ref{eq:A_PV}) receives contributions from $\gamma-Z$ and $\gamma-A_D$ interference terms, since the purely electromagnetic cross section does not contribute to the asymmetry. By calculating the relevant PVES asymmetries and the lepton charge asymmetry, we found that the total effect of the physical $Z$ and $A_D$ exchanges is given by the effective couplings~\cite{Thomas:2022qhj}
\begin{eqnarray}
\label{eq:C1q-C2q}
C_{1q} &=& C^Z_{1q} + \frac{Q^2 + M_Z^2}{Q^2 + M_{A_D}^2} C^{A_D}_{1q} \equiv C^{\rm SM}_{1q} ( 1 + R_{1q} ),\nonumber\\
C_{2q} &=& C^Z_{2q} + \frac{Q^2 + M_Z^2}{Q^2 + M_{A_D}^2} C^{A_D}_{2q} \equiv C^{\rm SM}_{2q} ( 1 + R_{2q} ) \,  ,\nonumber\\
C_{3q} &=& C^Z_{3q} + \frac{Q^2 + M_Z^2}{Q^2 + M_{A_D}^2} C^{A_D}_{3q} \equiv C^{\rm SM}_{3q} ( 1 + R_{3q} ) \,  ,
\end{eqnarray}
where $C^Z_q$ and $C^{A_D}_q$ have the same form as Eq.~(\ref{eq:C_q}), with $g_A$ and $g_V$ being replaced by the corresponding physical couplings in Eqs.~(\ref{eq:C_Z}-\ref{eq:C_AD}). $R_{1q}$, $R_{2q}$ and $R_{3q}$ characterise the corrections to the SM couplings, arising from the effects of a dark photon.

These correction factors depend on the dark parameters, $(\epsilon, m_{A_D})$, and the momentum transfer $Q^2$. We consider the region of interest (ROI) corresponding to $\epsilon \le 0.2$ in heavy mass region, which has not been fully excluded by the existing constraints. 

\begin{figure}[!h]
\begin{center}
\includegraphics[width=0.8\columnwidth]{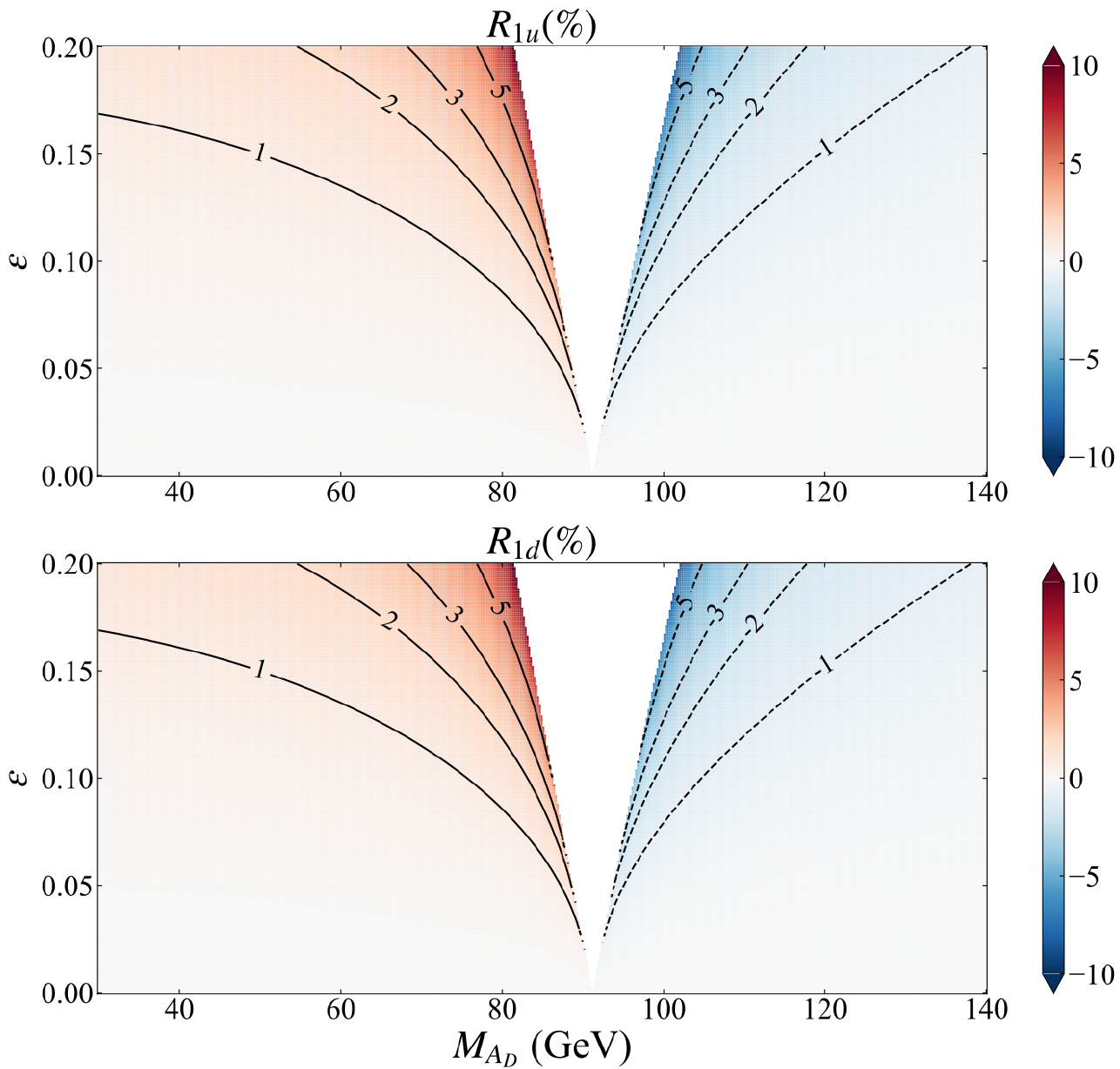}
%\vspace*{-0.2cm}
\caption{$R_{1u}$ and $R_{1d}$ at $Q^2 = 0.0045\ {\rm GeV}^2$. The gap is the ``eigenmass repulsion'' region in which the dark photon parameters are not accessible.}
\label{fig:R1q}
\end{center}
\end{figure}

At very low scale $Q^2 = 0.0045\ {\rm GeV}^2$ which is relevant for elastic scattering in the upcoming P2 experiment~\cite{Becker:2018ggl}, the corrections $R_{1q}$ are shown in Fig.~\ref{fig:R1q}. These could be as large as 5\% when the dark photon parameters approch to the ``eigenmass repulsion'' region. The corrections to the up and down quark couplings are roughly flavour independent and so cannot be simply represented by a change in the Weinberg angle~\cite{Thomas:2022qhj}.

\begin{figure}[!h]
\begin{center}
\includegraphics[width=0.8\columnwidth]{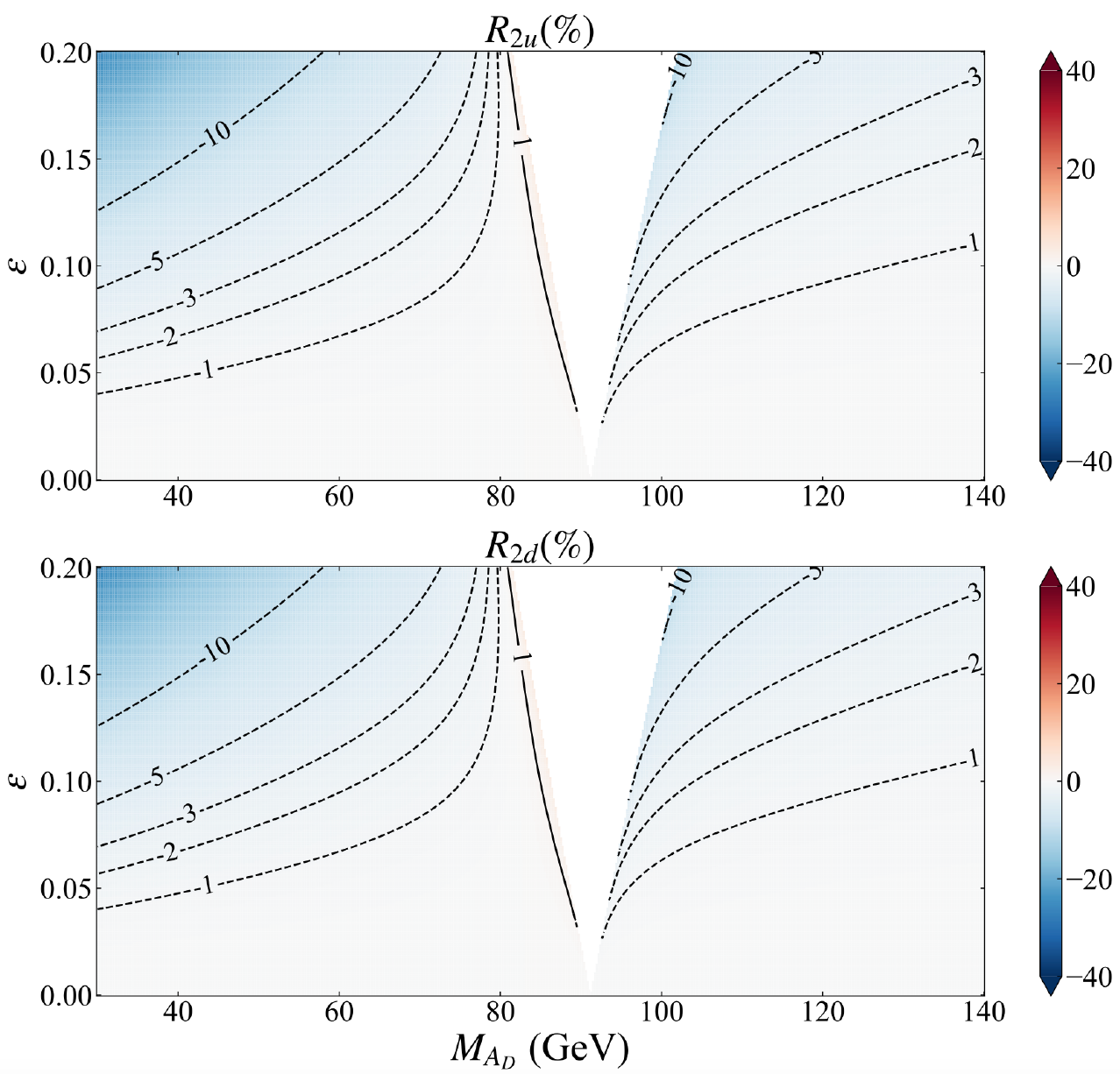}
%\vspace*{-0.2cm}
\caption{$R_{2u}$ and $R_{2d}$ at $Q^2 = 10^3\ {\rm GeV}^2$.}
\label{fig:R2q}
\end{center}
\end{figure}

At much higher momentum scale, $Q^2 =10^3\ {\rm GeV}^2$, accessible at HERA and EIC, the dark photon effects will lead to large corrections to the couplings $C_{2q}$ as shown in Fig.~\ref{fig:R2q}, while the corrections $R_{1q}$ are relatively small. $R_{2q}$ tend to be negative and as large as 10\%, suggesting large uncertainties in the extraction of valence parton distributions from high-$Q^2$ data.

\begin{figure}[!h]
\begin{center}
\includegraphics[width=0.8\columnwidth]{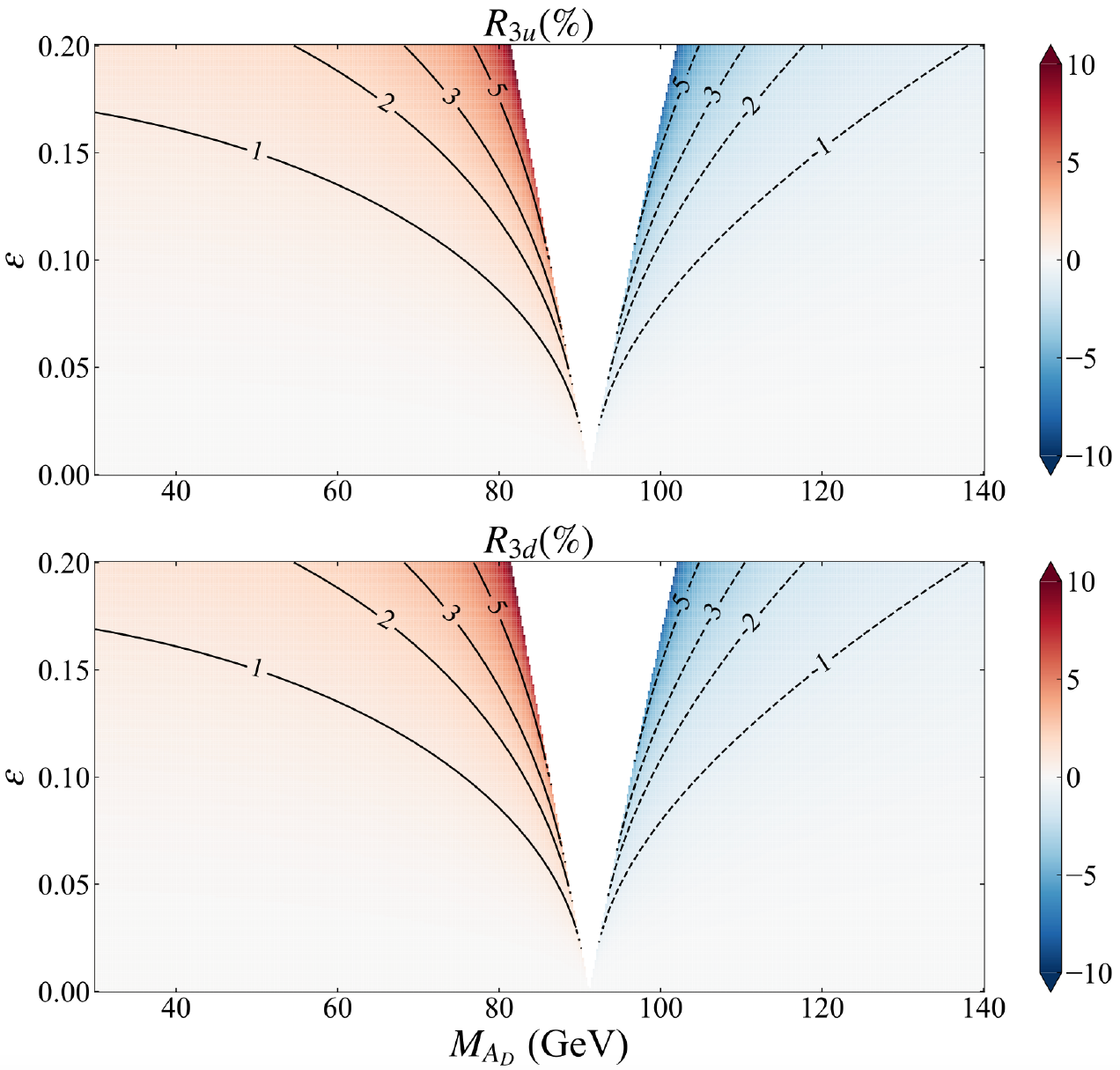}
%\vspace*{-0.2cm}
\caption{$R_{3u}$ and $R_{3d}$ at $Q^2 = 5\ {\rm GeV}^2$.}
\label{fig:R3q}
\end{center}
\end{figure}

Finally, Fig.~\ref{fig:R3q} shows the corrections to the couplings $C_{3q}$ at an average scale of $Q^2 = 5\ {\rm GeV}^2$ relevant for the planned SoLID experiment at JLab~\cite{JeffersonLabSoLID:2022iod}, which is expected to provide the first measurement of $C_{3q}$ in the future. We found that $C_{3q}$ would deviate from their SM predictions as large as 5\%, were a dark photon to exist.

\section{Favoured dark photon parameters}
\label{sec:fit}

\begin{figure}[!h]
\begin{center}
\includegraphics[width=0.9\columnwidth]{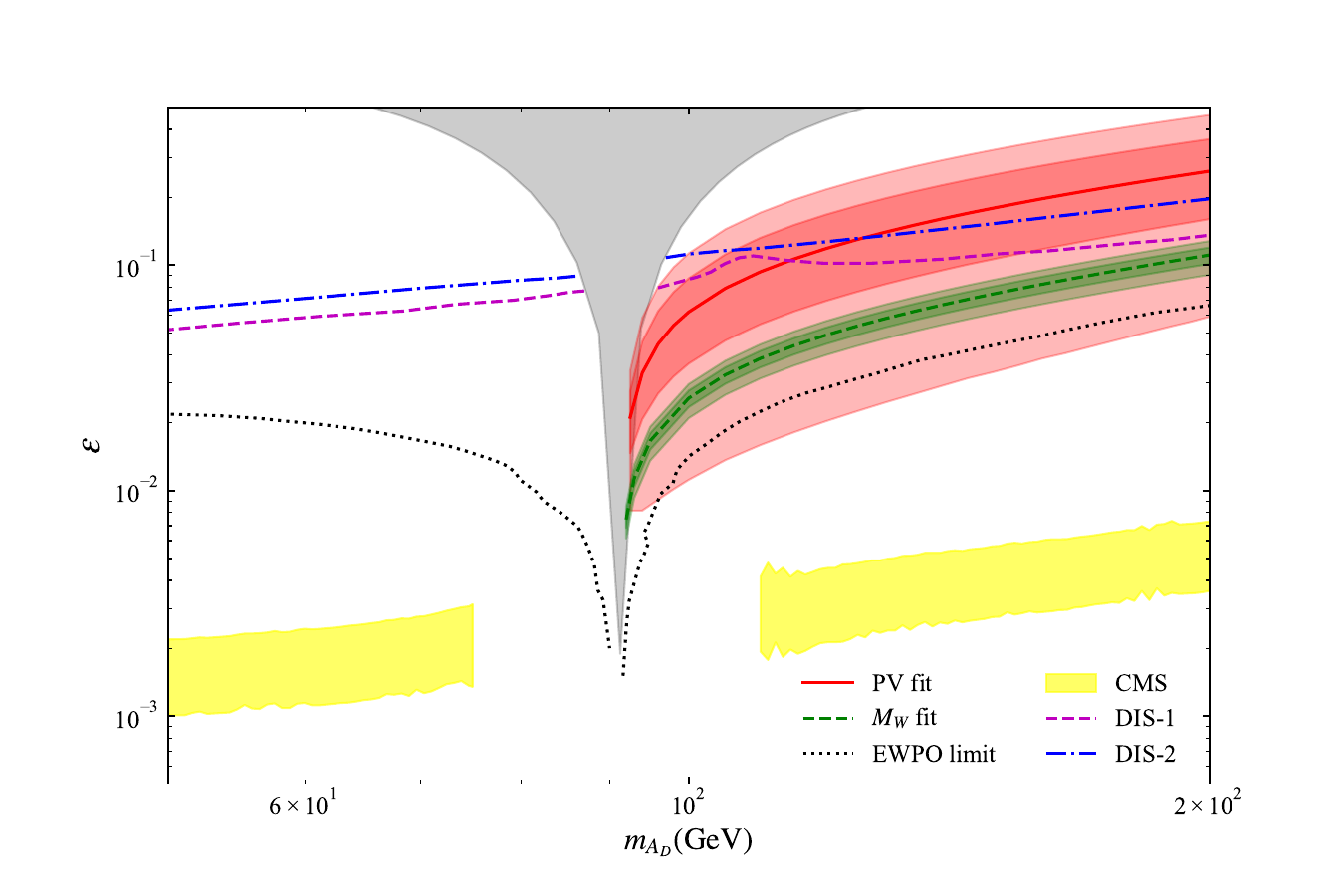}
%\vspace*{-0.2cm}
\caption{Favoured dark photon parameters from parity-violation data and the CDF $W$ boson mass, together with 68\% CL (dark band) and 95\% CL (light band) uncertainties~\cite{Thomas:2022gib}. The EWPO limit is taken from Ref.~\cite{Curtin:2014cca}. The DIS-1 and DIS-2 are taken from Ref.~\cite{Kribs:2020vyk} and Ref.~\cite{Thomas:2021lub}, respectively. We also show the 95\% CL exclusion constraints from the CMS Collaboration~\cite{CMS:2019buh}.}
\label{fig:fit}
\end{center}
\end{figure}

The currently available experimental data of PVES and atomic parity violation (APV) are summarised in Tab.~\ref{tab:PVES-DATA}.
The discrepancy between experiments and the SM predictions is characterised by $\chi^2_{\rm total} = 3.517$. We then perform a $\chi^2$ fit by including the dark photon~\cite{Thomas:2022gib}. The best values of $\epsilon$ are shown in Fig.~\ref{fig:fit} (red solid curve) for each value of $m_{A_D}$ above $m_Z$, corresponding to an improved $\chi^2_{\rm total} = 2.179$. In the region of $m_{A_D} < m_Z$, the inclusion of the dark photon will always worsen the $\chi^2$ with respect to the SM value.

The dark photon model can also be applied to explain the $W$ boson mass anomaly measured by the Collider Detector at Fermilab (CDF)~\cite{CDF:2022hxs}. The relation between the W-boson mass, $m_W$, and the Z-boson mass $m_{\bar{Z}}$ is~\cite{Awramik:2003rn},
\begin{equation}
m^2_W = m^2_{\bar Z} \left\{ \frac{1}{2} + \sqrt{\frac{1}{4} - \frac{\pi \alpha_{\rm em}}{\sqrt{2} G_F m^2_{\bar Z}} [1 + \Delta r(m_W, m_{\bar Z}, m_H, m_t,\ldots)]} \right\} \, ,
\end{equation}
where $\Delta r = 0.03677$ by taking $m_H = 125.14\ {\rm GeV}$ and $m_t = 172.89\ {\rm GeV}$.
Then the CDF result $m_W = 80.4335 \pm 0.0094\ {\rm GeV}$ implies $m_{\bar Z} = 91.2326 \pm 0.0076\ {\rm GeV}$. The dark photon parameters given by the green dashed curve in Fig.~\ref{fig:fit} are determined by shifting $m_{\bar Z}$ to the physical value $m_Z = 91.1875\ {\rm GeV}$ according to Eq.~(\ref{eq:m_Z_AD}), which also favour $m_{A_D} > m_Z$.

\begin{table*}[!htpb]
\renewcommand\arraystretch{1.2}
 \begin{center}
\begin{tabular}{ccccc}
\hline\hline
    {\rm Experiment}                          & $Q^2\ ({\rm GeV}^2)$ &                  {\rm data}                        &   SM       &   SM + dark photon  \\ \hline
Qweak~\cite{Qweak:2018tjf}                    &        $0.0248$      & $Q_{\rm w}^p = 0.0719 \pm 0.0045$                  &  $0.0708$ & $0.0707$ \\ 
PREX-II~\cite{PREX:2021umo, Corona:2021yfd}   &        $0.00616$     & $Q_{\rm w}(^{208}{\rm Pb}) = -114.4 \pm 2.6$        & $-117.9$  & $-117.1$ \\
PVDIS~\cite{PVDIS:2014cmd} $(\times 10^{-6})$ &        $1.085$       & $A^{{\rm exp}(1)}_{\rm PV} = -91.1 \pm 3.1 \pm 3.0$ &  $-87.7$  & $-87.2$  \\
                          \                   &        $1.901$       & $A^{{\rm exp}(2)}_{\rm PV} = -160.8 \pm 6.4 \pm 3.1$ & $-158.9$ & $-157.9$ \\ 
APV~\cite{ParticleDataGroup:2020ssz}          &           \          & $Q_{\rm w}(^{133}{\rm Cs}) = -72.82(42)$           & $-73.23$   & $-72.77$ \\ \hline \end{tabular}
\caption{The experimental data and the SM predictions of PVES and the atomic parity violation. The last column shows the fit results including the dark photon effects. }
\label{tab:PVES-DATA}
\end{center}
\end{table*}

\section{Conclusion}
\label{sec:conclusion}
We proposed a new tool for the dark photon searches through parity-violating electron scatterings.
We explored the sensitivity of PVES asymmetry to the dark photon parameters.
The dark photon effects are characterised by corrections to the SM couplings, which could be as large as 10\% for $C_{2q}$ at large $Q^2$,  and 5\% for $C_{1q}$ and $C_{3q}$ at low scales.

We also derived the dark photon parameters by fitting the parity-violation data and the CDF $W$-boson mass, favouring a heavy dark photon with mass above the $Z$-boson mass. The inclusion of a dark photon could improve the agreement between theory and a number of parity violation experiments.

The upcoming PVES experiments at P2~\cite{Becker:2018ggl}, SoLID~\cite{JeffersonLabSoLID:2022iod}, EIC~\cite{AbdulKhalek:2021gbh},  and MOLLER~\cite{MOLLER:2014iki} are expected to provide
more stringent constraints on the dark photon parameters. As one of the promising dark portal hypotheses, it is also essential to explore its implications in connection with dark matter particles.

\acknowledgments{
This work was supported by the University of Adelaide and the Australian Research Council through the Centre of Excellence for Dark Matter Particle Physics (CE200100008).
}

\bibliographystyle{JHEP}
\bibliography{bibliography}

\end{document}